\documentclass{article}
\usepackage[utf8]{inputenc}

\usepackage{authblk}
\usepackage{graphicx}
\usepackage[utf8]{inputenc}
\usepackage{amsmath}
\usepackage{amssymb}
\usepackage{xcolor}
\usepackage{appendix}
\usepackage{subcaption}
\usepackage{float}
\usepackage{booktabs}
\usepackage{physics}
\usepackage{nameref}
\usepackage{amsthm}
\usepackage[margin=1in]{geometry}
\usepackage[colorlinks=true, allcolors=blue]{hyperref}
\usepackage{letltxmacro}
\LetLtxMacro{\originaleqref}{\eqref}
\renewcommand{\eqref}{Eq.~\originaleqref}

\newcommand{\figkweek}{Figure \ref{fig:k_week_ahead}}
\newcommand{\figpid}{Figure \ref{fig:pid}}
\newcommand{\figpolicy}{Figure \ref{fig:param_comparison}}
\newcommand{\figasof}{Figure \ref{fig:as_of}}
\newcommand{\tabler}{Table \ref{tab:state_r2}}
\newcommand{\tableglobal}{Table \ref{tab:policy_corr}}
\newcommand{\tablestate}{Table \ref{tab:policy_corr_state}}
\newcommand{\branchingeq}{\eqref{eq:branching}}
\newcommand{\sireq}{\eqref{eq:sir_it}}
\newcommand{\closedloopeq}{\eqref{eq:closed_loop}}
\newcommand{\rhodefeq}{\eqref{eq:rho_def}}

\newtheorem{proposition}{Proposition}

\title{Current Implicit Policies May Not Eradicate COVID-19}

\author[a,b]{Ali Jadbabaie}
\author[a]{Arnab Sarker}
\author[a,c]{Devavrat Shah}
\affil[a]{Institute for Data, Systems, and Society, MIT}
\affil[b]{Department of Civil and Environmental Engineering, MIT}
\affil[c]{Department of Electrical Engineering and Computer Science, MIT}
\date{March 2022}

\begin{document}

\maketitle

\begin{abstract}
Successful predictive modeling of epidemics requires an understanding of the implicit feedback control strategies which are implemented by populations to modulate the spread of contagion.
While this task of capturing endogenous behavior can be achieved through intricate modeling assumptions, we find that a population's reaction to case counts can be described through a second order affine dynamical system with linear control which fits well to the data across different regions and times throughout the COVID-19 pandemic.
The model fits the data well both in and out of sample across the 50 states of the United States, with comparable $R^2$ scores to state of the art ensemble predictions.
In contrast to recent models of epidemics, rather than assuming that individuals directly control the contact rate which governs the spread of disease, we assume that individuals control the rate at which they vary their number of interactions, i.e. they control the derivative of the contact rate.
We propose an implicit feedback law for this control input and verify that it correlates with policies taken throughout the pandemic.
A key takeaway of the dynamical model is that the ``stable'' point of case counts is non-zero, i.e. COVID-19 will not be eradicated under the current collection of policies and strategies, and additional policies are needed to fully eradicate it quickly.
Hence, we suggest alternative implicit policies which focus on making interventions (such as vaccinations and mobility restrictions) a function of cumulative case counts, for which our results suggest a better possibility of eradicating COVID-19.
\end{abstract}

\section{Introduction}

Determining the optimal policy for regulating an epidemic requires assessing complex infection dynamics across heterogenous populations.
Due to this complexity, simplified models of epidemic spread are often used in order to develop guidelines and understand the impact of policies on the level of infection.
Throughout the COVID-19 pandemic, a wide range of models have been implemented in order to both provide forecasts for important time series related to the pandemic as well as estimate causal effects of various policies \cite{chernozhukov2021causal, ray2020ensemble}.
Such approaches have been developed since as early as the 19th century, and have been successfully applied to forecast epidemics such as seasonal influenza, smallpox, and H1N1 \cite{grassly2008mathematical}.

\begin{figure}[t]
    \centering
    \includegraphics[width=0.6\textwidth]{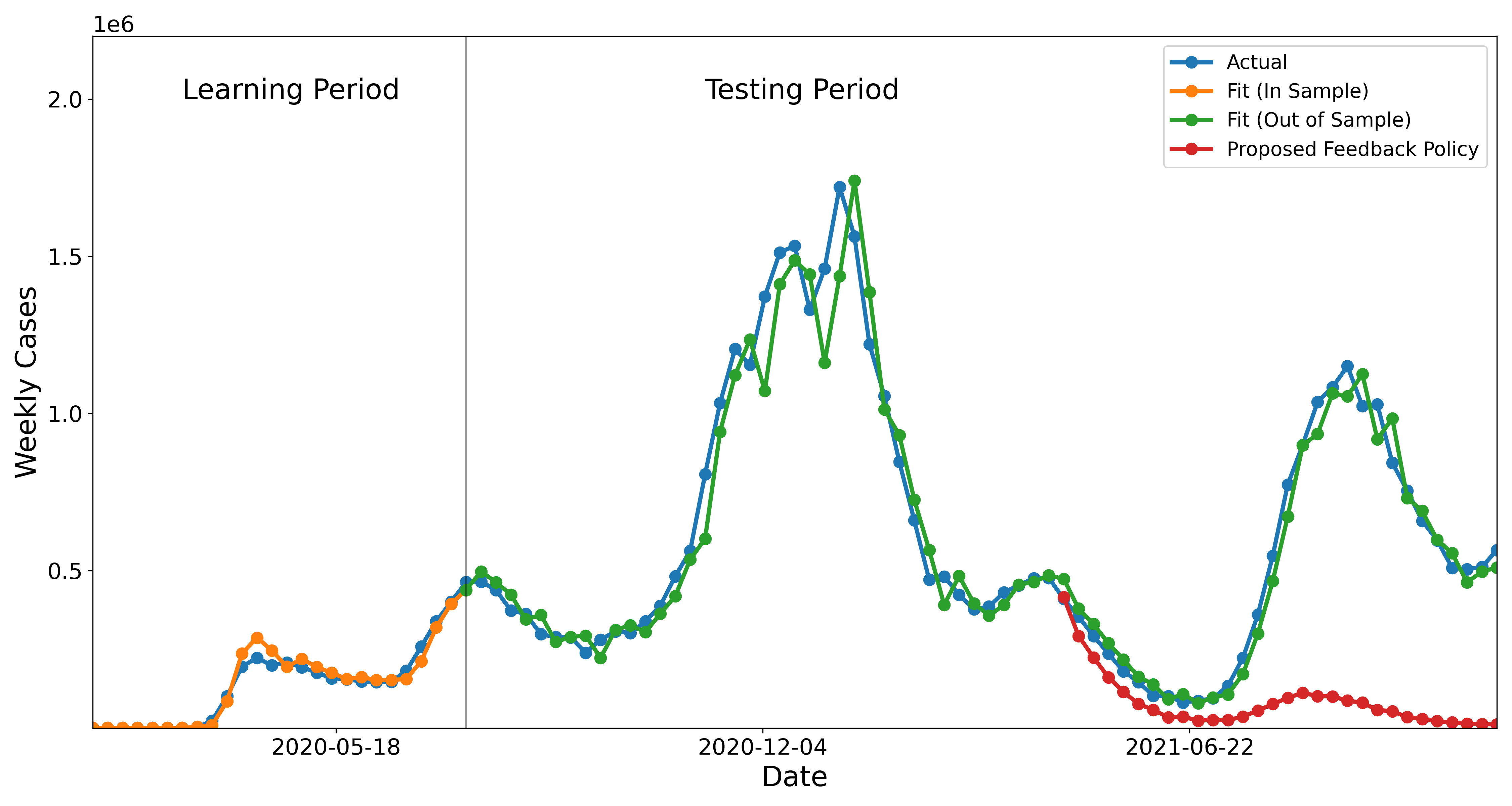}
    \caption{US Case counts as modeled by the dynamical system in \eqref{eq:closed_loop}.
    We learn the parameters of the control $\beta_1$ and $\beta_2$ on the first 30 weeks of data, and then they are held constant for the remainder of the data. 
    One week ahead predictions using this method are shown in orange (in-sample) and green (out of sample), and the true data is shown in blue.
    For the in-sample time series, the model fits the data with an $R^2$ value of $0.966$, and for the out of sample time series, the $R^2$ value is $0.949$.
    In comparison, the 1 week ahead forecasts of state of the art ensemble predictions over the same period is $0.942$ \cite{ray2020ensemble}.
    The red line represents a transition to a separate proposed policy which incorporates an integral feedback term, as discussed in \hyperref[ss:future]{Rethinking Control for Future Epidemics}.}
    \label{fig:us_fit}
\end{figure}

However, a key distinction between the COVID-19 pandemic and the spread of infectious disease in the past is the extent to which populations have reacted to limit the spread of the virus.
As cases have surged, government officials have instituted lockdowns and mask mandates, individuals have changed their behavior at an unprecedented scale, and the public has received access to mass vaccination.
This endogenous response to the state of the pandemic is rarely reflected by models developed for previous epidemics, but is critical in developing a robust response to COVID-19 \cite{stewart2020control, van2021pandemic}.
Even of the models used for COVID-19 death forecasting, only one explicitly includes an assumption that policy changes as the state of the pandemic worsens \cite{rahmandad2020risk}.
That is not to say that other models do not account for behavior.
Rather, the remaining predictive models which do account for behavior treat human intervention as an exogenous variable, observed for example through mobility data and the presence of government mandates and vaccinations.
In this work, we identify a model for epidemics where the response to the pandemic is encoded as an endogenous feature which depends on the magnitude of cases so far.

\subsection{Model Description}
We begin by considering a general model of the growth phase of an epidemic with a time-varying growth rate, which we show contains several common epidemic models as special cases.
We let $I(t)$ represent the number of infections in a population at time $t$, where $t$ is a discrete time index, and use the convention $I(0) = 1$.
Generically, any epidemic process with time-varying growth rate can be written in the following form:
\begin{equation}
\label{eq:branching_r}
    I(t+1) = I(t) \times \mathcal{R}(t) \times \eta(t) \,.
\end{equation}
Here, $\mathcal{R}(t)$ represents a generic time-varying parameter which indicates the number of infections that stem from each infected individual, and $\eta(t)$ represents noise in the process, which we assume to come from a log-normal distribution with parameters $\mu = 0$ and $\sigma^2$.
This choice of distribution for the noise is fundamental to the branching process literature that inspires this type of model and is a natural assumption in this setting~\cite{athreya1972branching}.

In this work, we build upon the aforementioned model and assume individuals control the growth rate using a time-varying parameter $\rho(t) \geq 0$ which represents the proportion of existing ties that each infected individual will add or remove at time $t$.
Mathematically, we assume $ \mathcal{R}(t) = \mathcal{R}(0) \times \prod_{k = 1}^t \rho(k)$, where $\mathcal{R}(0)$ is taken to be a constant value that represents the initial rate of infection.
Hence, the open loop dynamics of the system considered in this work take the form
\begin{equation} \label{eq:branching}
   I(t+1) = I(t) \times \mathcal{R}(0) \times \prod_{k = 1}^t \rho(k) \times \eta(t) \,.
\end{equation}
Although equating $\rho(t) = \mathcal{R}(t) / \mathcal{R}(t-1)$ reveals that the models in \eqref{eq:branching_r} and \eqref{eq:branching} are equivalent, we note that the parameterization of the model in terms of $\rho(t)$ represents a subtle but important distinction from models proposed in recent literature: rather than assume the contact rate is a direct function of a population's interventions, here we make the assumption that individuals are instead able to control the \emph{rate} at which they increase or decrease interactions with one another.
In terms of a mechanical analogy to a car, rather than assuming that individuals dictate control of position through a choice of velocity, our model assumes that individuals control their position through acceleration, i.e. by pressing on the gas or the brakes.
This parameterization allows us to model the reaction of the population in a novel way, by allowing $\rho(t)$ to be a function of previous case counts, as will be shown in \eqref{eq:rho_def}.

\paragraph{Connections to Existing Epidemic Models}
Because the time series of control parameters $\rho(t)$ is not specified in \eqref{eq:branching}, the model is considered open loop, and we note that natural selections of $\rho(t)$ lead to common epidemic models in the literature.
In particular, if $\rho(t) = 1$ for all $t$, then the model leads to \emph{exponential growth} in case counts, which is a commonality across the initial stages of many epidemic models \cite{hethcote2000mathematics}.

If $\rho(t) = \bar{\rho}$ for all $t$, where $\bar{\rho} < 1$, then \eqref{eq:branching} recovers the form of the Gaussian curve noted in \emph{Farr's law}, which is a common non-mechanistic approach to prediction of an epidemic \cite{farr1840progress}.

Moreover, a detailed connection can be made to the standard SIR model.
A specific time-varying choice of $\rho(t)$ can recreate the traditional \emph{Susceptible-Infected-Recovered (SIR)} model as well as its time-varying extensions \cite{kermack1927contribution}.
In discrete time, the SIR model consists of the following set of dynamic equations \cite{allen1994some}:

\begin{align*}
    S(t+1) - S(t) &= - \frac{\beta}{N} S(t) I(t) \\
    I(t+1) - I(t) &= \frac{\beta}{N} S(t) I(t) - \gamma I(t)\\
    R(t+1) - R(t) &= \gamma I(t) \,.
\end{align*}
In the model, $N$ represents population size, and $S(t)$, $I(t)$ and $R(t)$ represent the parts of the population that are susceptible, infected, and recovered from infection, respectively. 
In this simplified model, it is assumed that for all $t$, $S(t) + I(t) + R(t) = N$, and we use the convention that at $t = 0$ we have $I(0) = 1, R(0) = 0$, and $S(0) = N - 1$.
The parameters $\beta$ and $\gamma$ modulate the dynamics of the system, where $\beta$ measures the average number of contacts each individual in the population has with others, and $1/\gamma$ represents the average amount of time that an infectious individual is able to infect others.
Such parameters are often allowed to be time-varying, which ultimately results in infection dynamics of the form
\begin{equation}
    I(t+1) = I(t)  \times \left[ 1 - \gamma(t) + \frac{\beta(t)}{N} S(t) \right]
    \label{eq:sir_it} \,.
\end{equation}
For any choice of time-varying $\gamma(t)$ and $\beta(t)$, these dynamics are a special case of the model \eqref{eq:branching}, as we note in the following proposition, proved in the Supplementary Material.
\begin{proposition} \label{prop:equivalence}
There exists a parameterization of the open loop model in  \eqref{eq:branching} such that its dynamics are equivalent to the time-varying SIR model in \eqref{eq:sir_it}.
\end{proposition}
The connection to the SIR model reveals that the model in \eqref{eq:branching} provides a novel way to represent how individuals react in a pandemic.
Specifically, we see that the open loop model in \eqref{eq:branching} does not explicitly assume that individuals in a population are adjusting their contact rates or recovery rates.
The model assumes that people are unaware of the cardinal values that govern the physics of infection dynamics, given by a particular value of $\mathcal{R}(t)$, and instead control the relative proportion of ties they add or remove from their network, which would be equivalent to $\mathcal{R}(t) / \mathcal{R}(t-1)$.

\paragraph{A New Model of Implicit Feedback}
We allow the model in \eqref{eq:branching} to incorporate implicit feedback by allowing the control parameter to take the following form:
\begin{align}
    \log \rho(t) &= \beta_1 \times \log I(t)~+ \nonumber \\
    &\quad \beta_2 \times (\log I(t) - \log I(t-1) - \log \mathcal{R}(0)) + \beta_3 \label{eq:rho_def} \,.
\end{align}

Behaviorally, \eqref{eq:rho_def} implies that individuals in a population observe case counts through $\log I(t)$, and if $\beta_1 < 0$, which is the case in observed data, then larger magnitude of cases result in more control actions, for example through increased social distancing or mask adoption.
Similarly, individuals react to the recent change in cases relative to the reproductive rate ($\log I(t) - \log I(t-1) - \log \mathcal{R}(0)$), and when $\beta_2 < 0$, which is again consistent with empirical observations, then individuals increase the control action if cases are rising quickly.
Finally, $\beta_3$ represents a constant term which indicates an inherent level of reaction by the population.
It is worth noting that if $\beta_1 = \beta_2 = \beta_3 = 0$, then $\rho(k) = 1$ in this uncontrolled case and we recover a simple model of exponential growth.
Moreover, as discussed in the Supplementary Information, this assumption can be viewed as creating a specific time-varying structure for the time-varying SIR model in \eqref{eq:sir_it}.
Using state feedback allows us to write the model into the single closed loop dynamical system as follows.

Let $\mathbf{X}(t) \in \mathbb{R}^3$ represent the state of a dynamical system where $X_1(t) = \log I(t-1)$, $X_2(t) = \log I(t)$, and $X_3(t)$ represents the cumulative control taken so far, i.e. $X_3(t) = \sum_{k = 1}^t \log \rho(k)$.
The closed loop dynamics of the system \eqref{eq:branching} under the control policy \eqref{eq:rho_def} are summarized by the following affine dynamical system:
\begin{align}
    \mathbf{X}(t+1) &= \mathbf{Q} \mathbf{X}(t) + \mathbf{c} + \boldsymbol{\eta}(t) \,.\label{eq:closed_loop}
\end{align}
Here,
\begin{align*}
  \mathbf{Q} = \begin{bmatrix}
    0 & 1 & 0 \\
    0 & 1 & 1 \\
    - \beta_2  & \beta_1 + \beta_2 & 1
    \end{bmatrix} \,, ~\text{and}~\mathbf{c} = \begin{bmatrix}
    0\\ \log \mathcal{R}(0)  \\ -\beta_2 \log \mathcal{R}(0) + \beta_3
    \end{bmatrix} \,.
\end{align*}
Given the assumption that $\eta(t)$ in \eqref{eq:branching} is log-normal, we see that $\boldsymbol{\eta}(t)$ can be modeled as unobserved Gaussian noise with mean $0$ and variance $\sigma^2$.
This closed loop model is extremely simple as it is governed by only three parameters, and can be efficiently learned from data \cite{sarkar2019near}.
As our results indicate, the model fits the data surprisingly well, which suggests that it succinctly describes existing behavior throughout the pandemic, and the model has the critical implication that steady state infections are non-zero.

\begin{figure}
    \centering
    \includegraphics[width=0.6\textwidth]{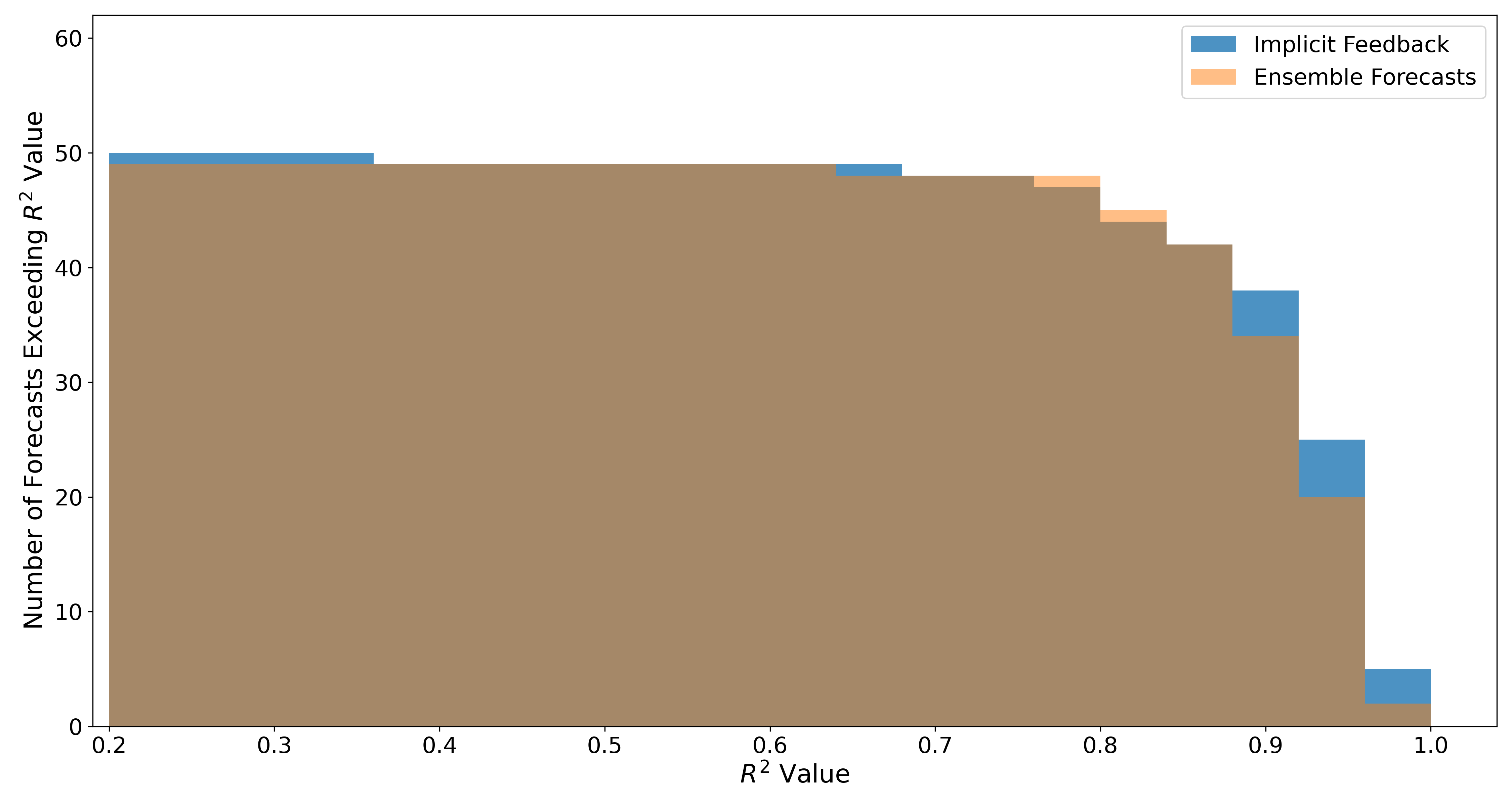}
    \caption{Comparison of $R^2$ values between the implicit feedback approach in \eqref{eq:closed_loop} to ensemble forecasts \cite{ray2020ensemble} on each state.
    The plot shows the number of states for which each method has an $R^2$ at least that of the value on the $x$-axis.
    Calculated $R^2$ values can be found for each state in  \tabler~of the Supplementary Information.
    Even though the closed model in \eqref{eq:closed_loop} only uses 3 parameters learned in a training period, whereas the ensemble method is aggregated over multiple complex models each week, we find that the closed loop model outperforms the ensemble, particularly in the region where $R^2 > 0.85$, as is shown by the blue bars exceeding the orange.
    In contrast, a fixed parameter SEIR model, which also has three parameters, does not achieve positive $R^2$ for one week ahead forecasts on any state in the dataset.
    }
    \label{fig:state_r2_comp}
\end{figure}

\subsection{Summary of Results} With the above model, in comparing to existing data we see that there is a satisfactory fit to observed data and that the parameters of the implicit feedback correlate with observed policies.
Moreover, in our analysis, we find that a major consequence of this model is that weekly cases are never eradicated; rather, for most observed learned parameters, weekly cases will stabilize at some non-zero level.
This has non-trivial policy implications-- in order to eradicate COVID-19 sooner rather than later, populations need to introduce another policy intervention. 
For example, by reacting to \emph{total} case counts since the beginning of the pandemic rather than the case counts in the last two weeks.

\paragraph{Fit to Data.} Our results indicate the model fits to empirical data surprisingly well, despite having only 3 learned parameters. 
As a means of comparison, we assess the performance of our model against the ensemble method used by the CDC, which aggregates predictions from at least 30 state of the art forecasting models each week to provide predictions of COVID-19 case counts \cite{ray2020ensemble}.
In Figure \ref{fig:state_r2_comp}, we see that the $R^2$ is comparable to state of the art ensemble predictions across all 50 states, when the method is trained using 65 weeks of data.
In fact, the approach outperforms the CDC point forecasts in 32 states.
The implicit feedback approach also outperforms a Susceptible-Exposed-Infected-Recovered (SEIR) model, which similarly has three parameters but does not achieve a positive $R^2$ score on any state.
An example of the one week ahead forecasts can also be seen in Figure \ref{fig:us_fit}, where we see similar results across data for the entire United States.
We also show that other similar feedback policies do not fit as well to the data as the simple policy suggested in \eqref{eq:rho_def}.
This, combined with the comparison to observed policies noted below, gives additional credence to this choice of model.

\paragraph{Comparison to Observed Policies.} In regressing parameters of the model against policies taken by different populations across the globe, we find significant correlations between the $\beta_1$ and $\beta_2$ parameters when compared to mobility data as well as levels of natural immunity measured by the percentage of the population which has had a confirmed positive COVID-19 test.
This relationship between parameters of the model and observed policies taken by populations reinforces the validity of the model and also suggests ways in which the effect of the pandemic may be mitigated which are consistent with the prevailing public health guidance in the United States.

\paragraph{Impossibility of Eradication.}  We show that there is no selection of $\beta_1,$ $\beta_2$, and $\beta_3$ that has been implemented by states which would result in zero weekly cases.
As a result, in order to fully eliminate the COVID-19 pandemic, populations must alter their implicit strategies and deviate from the dynamical model in \eqref{eq:closed_loop}.
This result is consistent with previous results on the fragility of non-pharmaceutical interventions, which suggests that even small measurement errors in the timing of a strategy can produce large increases in case counts \cite{morris2021optimal}.
Our results expand on this prior work by incorporating recent data, suggesting that even pharmaceutical interventions such as vaccinations are insufficient to eradicate new COVID-19 cases. 
Further, the results presented here are empirical in nature, and highlight properties of implemented policies.

We then conclude by presenting an example of a policy which does result in the eradication of COVID-19 cases in theory, and provide steps which suggest a time-varying schedule of vaccinations and stay at home orders which would approximate such a policy.
Our results suggest that by increasing the intensity of certain policies by an order of magnitude, COVID-19 cases can be eradicated.

\section{Data}
The data used in order to learn the parameters of the model is taken from Johns Hopkins \cite{dong2020jhudata}.
In preprocessing the data, we assume each time index $t$ represents a full week to remove weekly seasonality.
The dataset provides time series of case data in each state of the United States, as well as for 110 countries, which we aggregate weekly to fit the model in \eqref{eq:closed_loop}.
We also use this data to compute estimates of naturalized immunity, by taking the ratio of confirmed positive tests and each region's population to get a percentage of individuals who have had COVID-19.

In order to compare the learned parameters to observed policies, we also use data from Google \cite{aktay2020google}, which compares aggregate mobility trends across individual states and countries compared to pre-pandemic levels.

\section{Methods}
\begin{figure}
    \centering
    \includegraphics[width=0.6\textwidth]{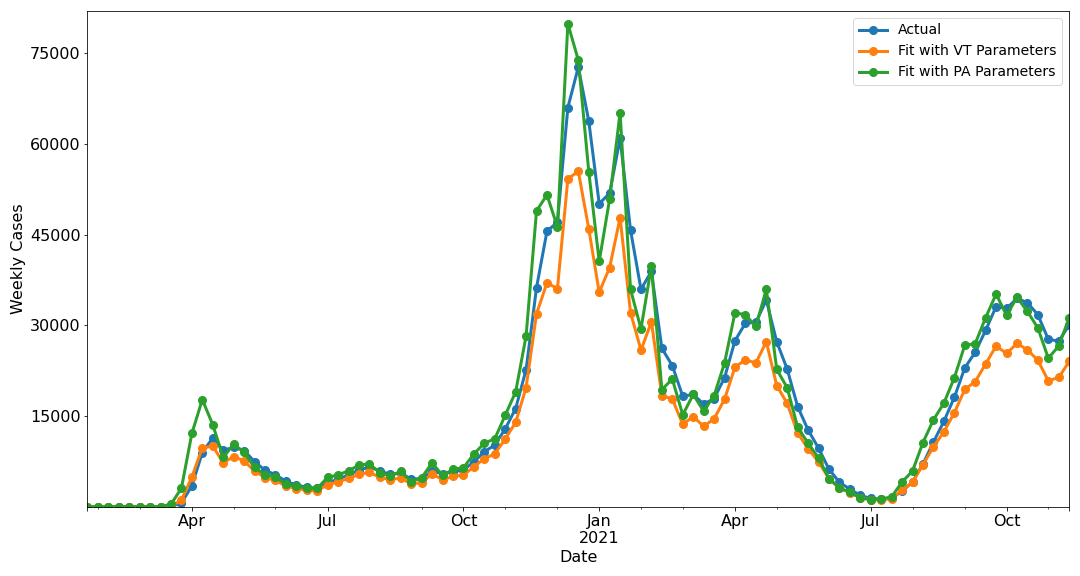}
    \caption{Applying learned parameters across different states. We find that different values of the implicit control policy correspond to dramatically different observations in case counts, suggesting that variations in the control parameters $\beta_1$, $\beta_2$, and $\beta_3$ are meaningful.}
    \label{fig:pa_vt}
\end{figure}

To learn the parameters of the implicit control, we focus on identifying the parameters $\beta_1$, $\beta_2$, and $\beta_3$ from \eqref{eq:rho_def}.
To do so, we first infer each $\log \rho(t)$ from future time steps, as \eqref{eq:branching} yields:
\[ \log \rho(t) = \log I(t+1) - \log I(t) - \sum_{k = 1}^{t-1} \log \rho_k - \log \mathcal{R}(0)   - \log \eta(t) \,.\]
Hence, because $\log \eta(t)$ is mean-zero, we estimate
\begin{align*}
    \log \rho_0 &= \log I(t+1) - \log I(t) - \log \mathcal{R}(0) \\
    \log \rho(t) &= \log I(t+1) - \log I(t) - \sum_{k = 1}^{t-1} \log \rho_k - \log \mathcal{R}(0) ~~ (t \geq 1) \,.
\end{align*}
We estimate $\mathcal{R}(0) $ to be equal to 2.5 \cite{linka2020reproduction}, and hence we are able to infer the values of the dependent variable $\log \rho(t)$.
Next, we estimate the regressors on the right hand side of \eqref{eq:closed_loop}, namely $\log I(t)$ and $\log I(t) - \log I(t-1) - \log \mathcal{R}(0) $.
We then perform a least-squares regression using the inferred $\log \rho(t)$ values and the observed regressors to determine the coefficients $\beta_1$, $\beta_2$, and $\beta_3$.
Such an approach of using a least-squares regressions to learn the parameters of a dynamical system are not uncommon, and many theoretical results exist showing the consistency and non-asymptotic reliability of such methods \cite{sarkar2019near}.

We also provide statistical evidence which shows that the learned $\beta_1$ and $\beta_2$ parameters above correlate with policies taken by individuals during the pandemic.
To get an estimate of policies taken during the pandemic, we process the Google mobility data, which consists of six mobility measures given as time series for each geographical region, and take the first two principal components of the data, such that for each region we get two time series.
We then average this time series for each region over the time period that the policies are learned, giving two regressors for each region.
Finally, we compute the correlation between the $\beta_1$ and $\beta_2$ parameter and each mobility metric.
We then repeat the same procedure replacing the mobility metrics with the normalized number of cumulative cases in each region up to the end of the training period, which we use as a proxy for natural immunity.

\section{Results}
\begin{figure}
    \centering
    \includegraphics[width=0.6\textwidth]{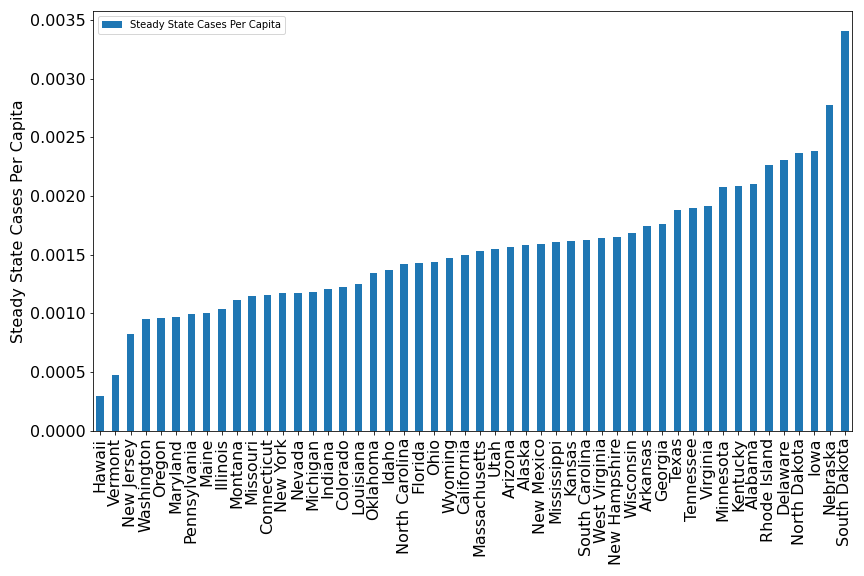}
    \caption{Computed steady state case counts per capita for each state, based on the learned parameters of the implicit control policy. 
    These values have a 0.564 correlation with the actual mean weekly case counts per capita in each state $(p < 10^{-4})$, supporting the validity of the model \eqref{eq:closed_loop} and suggesting particular geographic regions to focus on adjusting implicit control.}
    \label{fig:stability}
\end{figure}

\subsection{Fit to Observed Data}

As shown in Figures \ref{fig:us_fit} and \ref{fig:state_r2_comp}, the use of the closed-loop model \eqref{eq:closed_loop} fits the data well, with $R^2$ values comparable to the 1 week ahead CDC ensemble forecasts which aggregate predictions from over 30 state of the art prediction models \cite{ray2020ensemble}.
It is worth noting that the CDC ensemble forecasts are used solely as a baseline to validate that the model in \eqref{eq:closed_loop} fits well to the data.
In practice, due to delays in the data collection, the observed values of $I(t)$ at time $t$ are often underestimated, resulting in a small drop in predictive performance when only data available at each particular time is used (Supplementary Information, \figasof).
Although such data fidelity issues can be accounted for, we note that the purpose of this work is to highlight the behavioral implications of this model as opposed to its predictive value.

\begin{figure*}[t]
    \centering
    \includegraphics[width=0.8\textwidth]{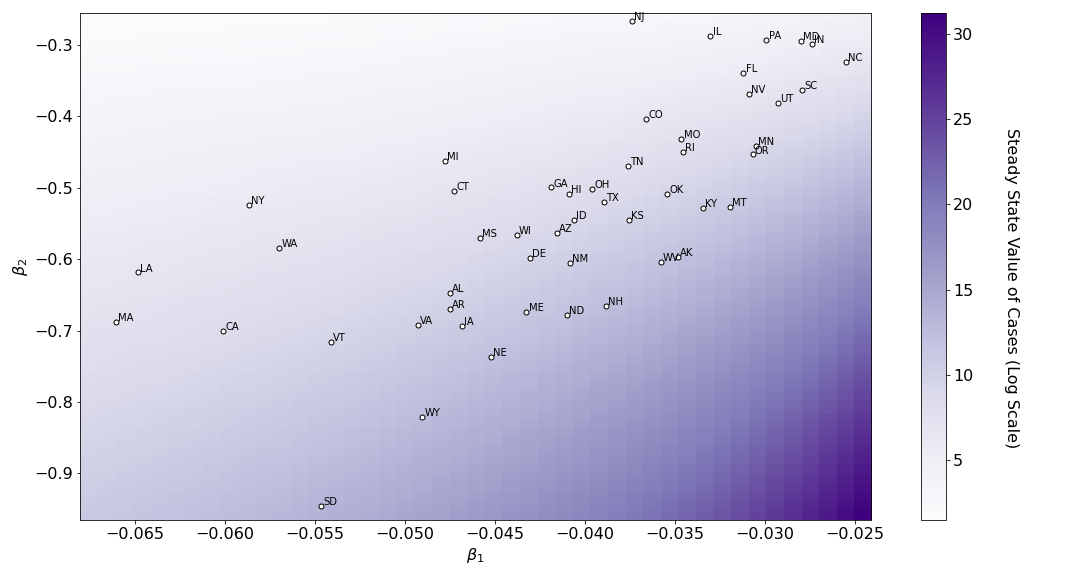}
    \caption{Coefficients of the implicit control used by the 50 US states. We find that states with low magnitude of $\beta_2$ and high magnitude of $\beta_1$ have a higher magnitude of steady state cases, highlighting a need for adjusted implicit policies in states such as South Dakota, Wyoming, and North Dakota.}
    \label{fig:state_coefs}
\end{figure*}

The learned $\beta_1, \beta_2$, and $\beta_3$ parameters are heterogenous between different geographic regions, and we find that different parameters of implicit control result in different implications for the pandemic.
The 50 states in the US provide insight into the different values of $\beta_1$ and $\beta_2$ and their importance in understanding the different forms of control that are being implicitly used by different states throughout the pandemic. 
The learned $\beta_1$ and $\beta_2$ parameters from each of the 50 states are shown in Figure \ref{fig:state_coefs}, and it is clear that there is heterogeneity in the learned values.

The differences in these learned values are significant as far as their effects on prediction, as we can not simply take one state's parameters and use this information to predict for a different state (Figure \ref{fig:pa_vt}). 
While there are clusters of states with similar parameters, it is clear that the specific values of parameters is still important in terms of generating valid predictions and explaining behavior.

The learned $\beta_1$, $\beta_2$, and $\beta_3$ parameters can also be used in order to understand the heterogeneity in each state's implicit control policies. 
When the matrix $\mathbf{Q}$ is stable in \eqref{eq:closed_loop}, we can compute the steady state number of cases $\lim_{t \rightarrow \infty} X_2(t)$. We report the per capita results in Figure \ref{fig:stability}.
This steady state value provides insight on how different states handled the pandemic in terms of the ability to efficiently bring case counts towards 0.
In particular, larger steady state values seem to suggest relaxed policies in controlling pandemic. 

It is worth noting that this is a compounded impact of government policies, population behavior and other circumstances (e.g. weather, industry, etc.). 
However, it is a definitive way to ``evaluate'' the compounded effect across states which may be of interest in its own right.

\subsection{Relationship to Observed Policies}
Our results indicate that both $\beta_1$ and $\beta_2$ correlate with observed policies taken throughout the pandemic.
In comparing the learned parameters of all 50 states as well as 110 countries to observed mobility patterns, we find statistically significant correlations (Supplementary Information, \tableglobal).
Moreover, we also see significant correlations with respect to the proportion of the population that has become immune during the training period.
This suggests a relationship between learned parameters and the proportion of the population which is no longer susceptible to the disease.
In fact, a similar relationship can be shown between $\beta_1$ and vaccination rates across the United States; however, the statistical evidence only suggests a correlation as the parameters are learned from the initial stages of the pandemic, so a causal relationship can not be made (Supplementary Information, \tablestate).

\begin{figure*}[t]
    \centering
    \begin{subfigure}[b]{0.47\textwidth}
        \includegraphics[width=\textwidth]{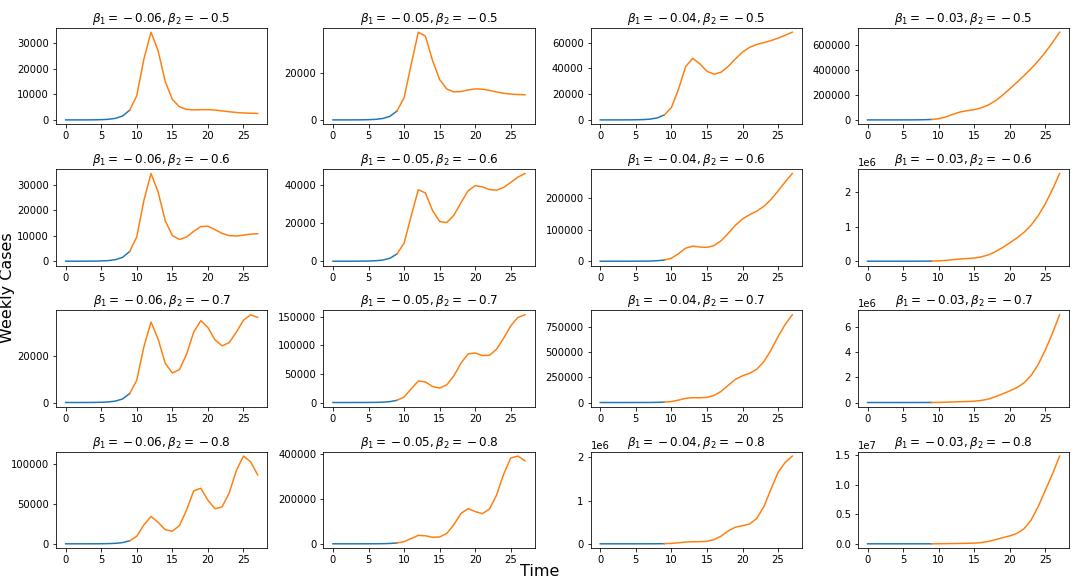}
        \caption{Simulated policy with control consistent with population behavior (\eqref{eq:rho_def})}
        \label{fig:beta_sweep}
    \end{subfigure}
    \hfill
    \begin{subfigure}[b]{0.47\textwidth}
        \centering
        \includegraphics[width=\textwidth]{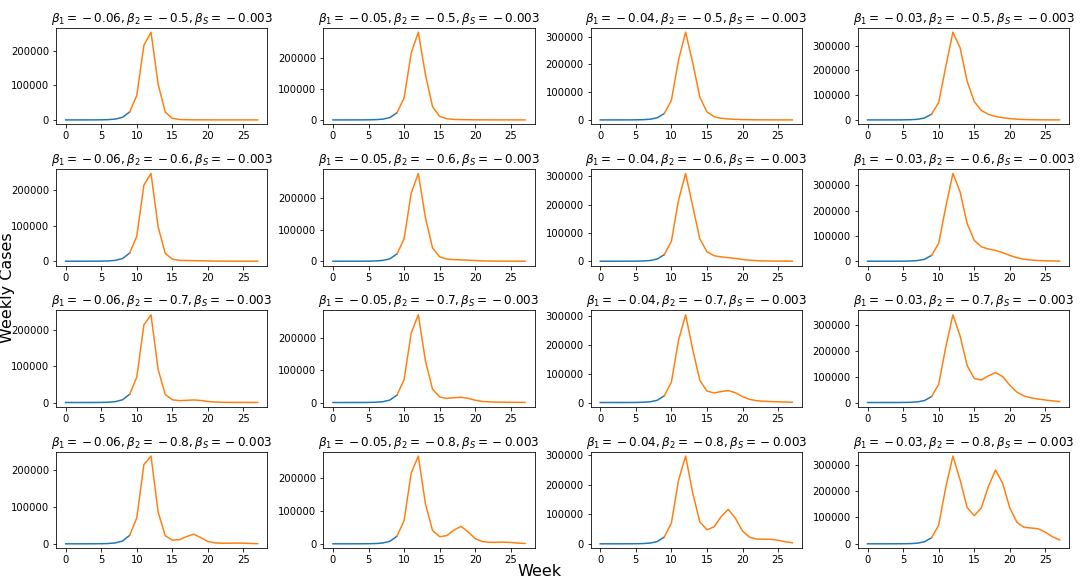}
        \caption{Simulated policy with an added integral term (\eqref{eq:rho_def_integrator})}
        \label{fig:beta_sweep_integrator}
    \end{subfigure}
    \caption{(a) Different choices of $\beta_1$ and $\beta_2$ for the system in  \eqref{eq:closed_loop}. This simulated system begins with an initial exponential growth (blue) followed by case counts when the control policy enacted (orange).
    We find that as $\beta_1$ becomes closer to 0, the system becomes increasingly unstable and prone to exponential growth. 
    As $\beta_2$ decreases away from 0 with fixed $\beta_1$, we see that the system becomes more oscillatory and likely to become unstable as well.
    Moreover, across all selection of parameters, even when the system is such that $\mathbf{Q}$ is stable, we see that the steady state number of cases is non-zero. (b) Adding an integrator term $\beta_S \sum_{k=1}^t X(k)$ to the implementation of the control input $\log \rho_t$. The simulated system again begins with an initial exponential growth (blue) followed by case counts when the control policy enacted (orange). 
    The integral term stabilizes the system and brings weekly case counts towards an equilibrium with no weekly cases, although in practice we find that no country has a statistically significant non-zero $\beta_S$ term in its control policy.}
\end{figure*}

\subsection{Comparisons to Other Feedback Models}

While the model fits to empirical data suggest that the model is consistent with observations, and the parameters do correlate with observed policies, such results cannot definitively prove that the model assumptions are representative of reality. 
Here, we compare our model to other plausible hypotheses in order to understand the value and limitations of this simple approach.

Because the results until now have provided one week ahead forecasts, and hence are dependent on most recent actual observations, it is also important to compare against a case in which these recent observations are estimated. 
In control theory, an observer is used for this task \cite{aastrom2010feedback}. 
When implementing the system with an observer as opposed to true data on the US case counts, we find that our predictions are still very close to when the original data is used, with nearly equivalent $R^2$ values. 
This approach also provides a way to produce $k$-week ahead forecasts for $k > 1$, and we find that short term forecasts which predict 1 or 2 weeks ahead are comparable to the CDC ensemble, whereas longer term forecasts deteriorate rapidly (Supplementary Information, \figkweek). 

We also compare this model to a simple, one-dimensional Proportional-Integral-Derivative (PID) controller, since PID control is nearly ubiquitous in control applications \cite{desborough2002increasing}. 
In this comparison, we find that the PID controller requires more data in order to effectively learn appropriate parameters, and that it does not explain the data as well as the closed-loop system in \eqref{eq:closed_loop} (Supplementary Information, \figpid). 
Hence, although PID control is omnipresent in engineering applications, it is not representative of the actual behavior observed throughout the pandemic.

\begin{figure}
    \centering
    \includegraphics[width=0.6\textwidth]{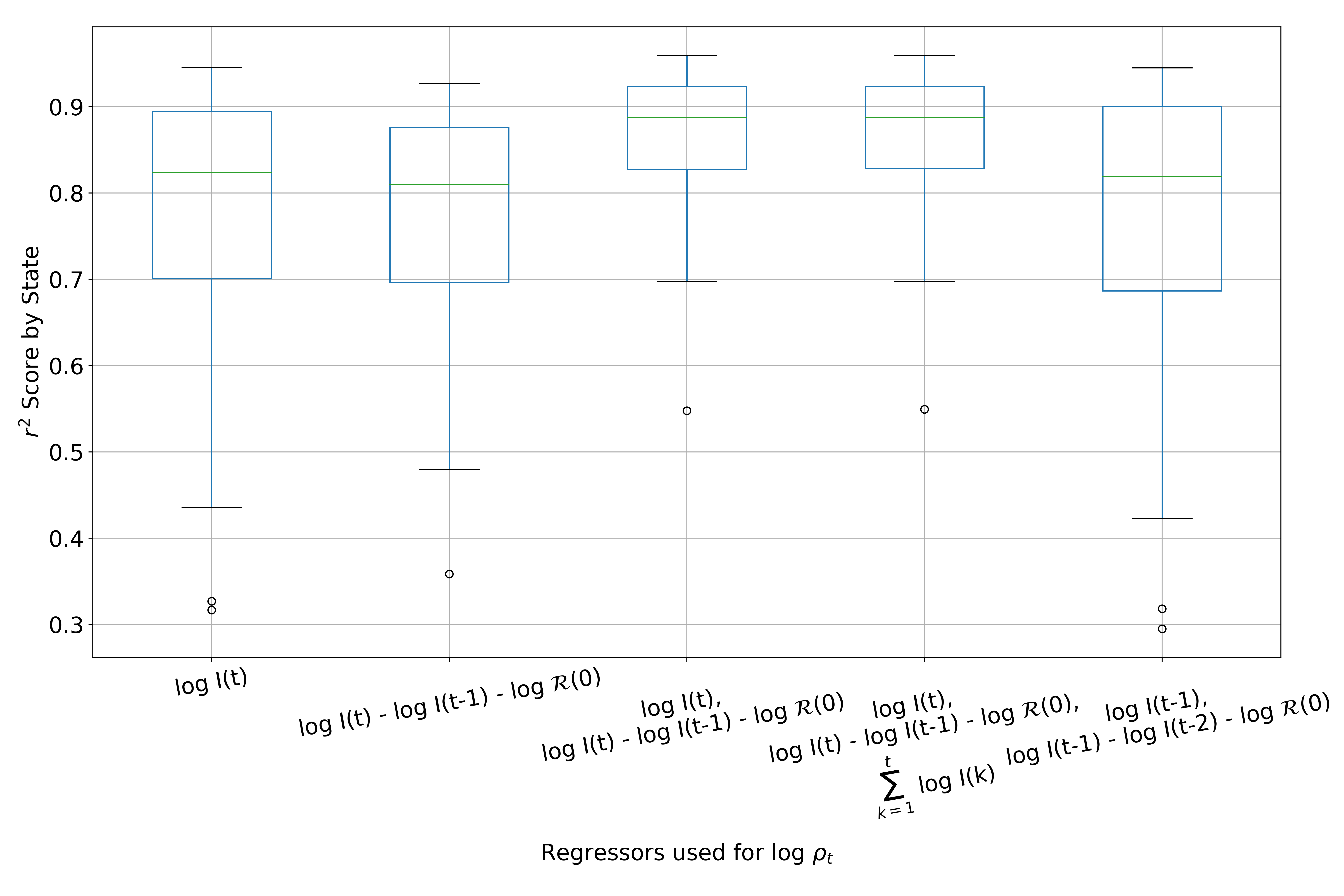}
    \caption{Boxplot of $R^2$ scores of the 50 US states when different regressors are used to determine $\log \rho_t$. We find that the choice of regressors in \eqref{eq:rho_def} provides the best fit to the data overall, suggesting that the three parameter control input fits sufficiently well across heterogenous regions.
    Moreover, including a regression term on $\sum_{k = 1}^t \log I(t)$, as is proposed in \eqref{eq:rho_def_integrator} does not result in an improved fit to the data, suggesting that this particular policy which would eradicate COVID-19 cases is not being implemented by populations. }
    \label{fig:state_r2}
\end{figure}

Finally, we compare against several models for which different sets of regressors are used in estimating $\log \rho_t$, in order to understand the different implications of each approach (Figure \ref{fig:state_r2}). 
Our results indicate that our proposed feedback law in \eqref{eq:rho_def} provides the best explanation of $\log \rho_t$ across the 50 states. 
It is also important to note that the use of $\sum_{k = 1}^t \log I(t)$ as a regressor for $\log \rho_t$ does not improve prediction performance, and that the coefficient of $\sum_{k = 1}^t \log I(t)$ is never significantly away from 0 across all 50 states and countries considered. 
This observation indicates at least one way in which policy changes can be modified to eradicate COVID-19 cases, and will be discussed further in \hyperref[ss:future]{Rethinking Control for Future Epidemics} where we discuss the policy implications of our results.

\subsection{Properties of the Implicit Feedback Control}

The impact of the parameters which govern the implicit feedback control is summarized in Figure \ref{fig:beta_sweep}. 
Figure \ref{fig:beta_sweep} indicates that there is a complex relationship between the values of $\beta_1$ and $\beta_2$ and the resulting case counts. 
Namely, increases in $\beta_2$ often result in oscillatory behavior, but still depend on the value of $\beta_1$. 
These plots reflect that an equilibrium for this type of control strategy occurs at a non-zero but constant steady state of weekly cases. 
Stated differently, the equilibrium of this approach suggests that behavior eventually results in an instantaneous reproductive rate of $1$.
This reinforces that the system is sensitive to the selection of $\beta_1$ and $\beta_2$, and our results indicate that $\beta_1$ and $\beta_2$ can be learned from data and fit the data surprisingly well (Figure \ref{fig:us_fit}).

\subsection{Rethinking Control for Future Epidemics}
\label{ss:future}

Principles of control dictate that when \eqref{eq:branching} has a multiplicative factor of $\mathcal{R}(0) > 1$, the system generated when taking logarithms of \eqref{eq:branching} can be driven to 0 weekly cases using a feedback law known as a double integrator. 
That is, rather than the system dynamics being governed by the form of $\log \rho_t$ described in \eqref{eq:rho_def}, a suitable control would take the form
\begin{align}
    \log \rho(t) &= \beta_1 \times \log I(t)~+ \nonumber \\
    &\quad \beta_2 \times (\log I(t) - \log I(t-1) - \log \mathcal{R}(0)) ~+\nonumber \\
    &\quad \beta_3 + \beta_S \times \sum_{k = 1}^t \log I(t)
    \label{eq:rho_def_integrator} \,,
\end{align}
where $\sum_{k = 1}^t \log I(t)$ represents an integral feedback control.
On synthetic data, the effect of this integrator term $\beta_S \sum_{k=1}^t \log I(t)$ is clear, as it drives weekly case counts to 0 (Figure \ref{fig:beta_sweep_integrator}). 
This additional term forces the weekly reaction of the community to the pandemic to go from being a proportional derivative (PD) control in \eqref{eq:rho_def} to a proportional-integral-derivative (PID) control, which provides the necessary integration of error which eventually results in a steady state of 0 weekly cases.

{\bf Key policy implication.} Ultimately, this suggests that the \emph{cumulative} costs of the pandemic should be emphasized, even when the current state of the pandemic has a low number of cases.
When the implemented control has little memory of the past, we can not expect cases to decay towards 0.
The policy implication here is \emph{implicit}, rather than explicit, in the sense that the current policy interventions being used, such as stay at home orders, mask mandates, and vaccination development, are still suggested by this model.
However, rather than implementing these policies a function of recent cases, the model suggests they should be implemented and intensified according to cumulative case counts.
Moreover, because the suggestion is implicit in behavior, the model also calls for individuals in a population to adopt a slightly different latent approach that reduces contacts as a function of cumulative case counts as opposed to recent ones.

\begin{figure}
    \centering
    \includegraphics[width=0.6\textwidth]{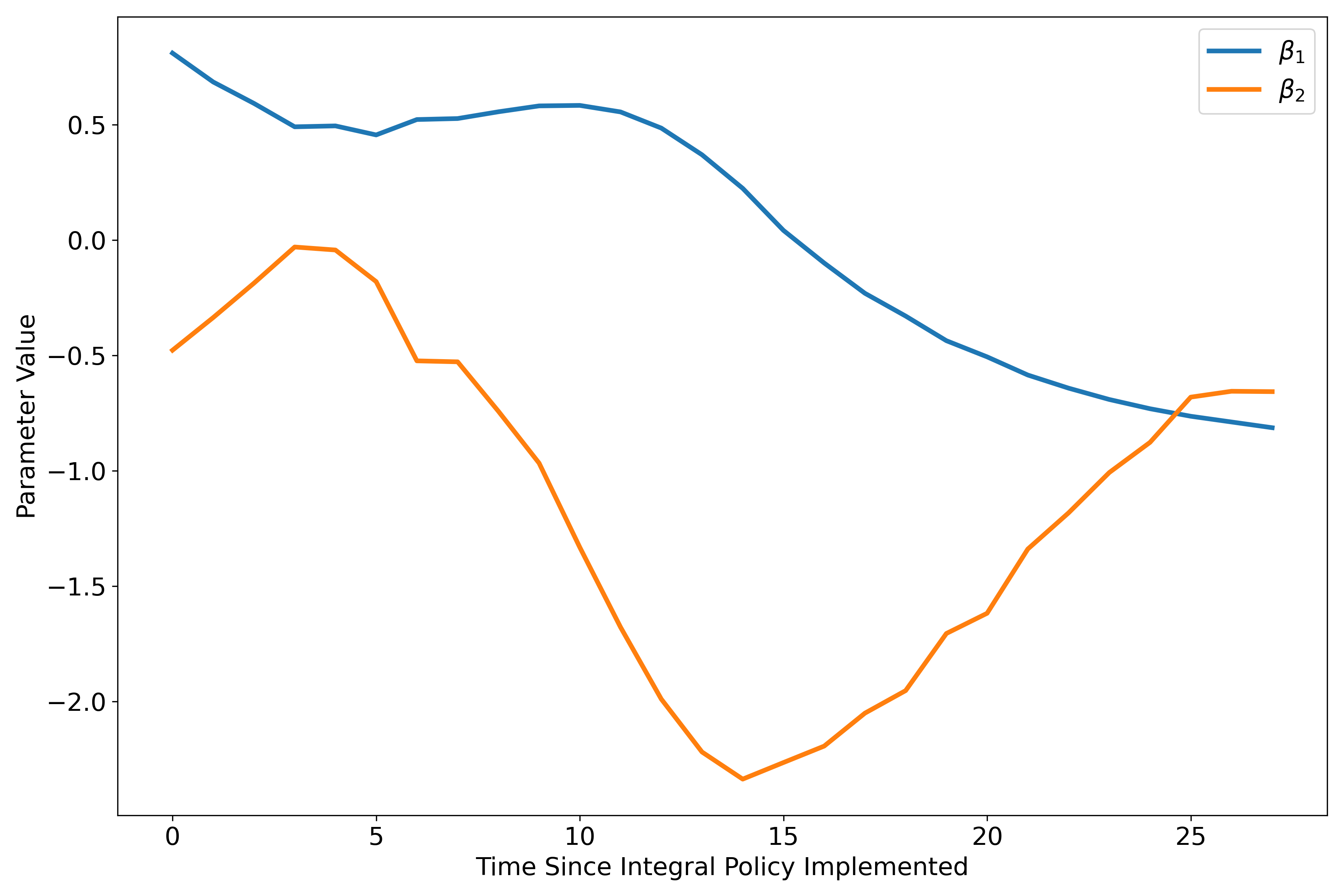}
    \caption{Example of recreating the integral feedback policy of the red line in Figure \ref{fig:us_fit} using time-varying $\beta_1$ and $\beta_2$ to replicate the effect of an integral feedback term.
    By steadily increasing vaccination adoption according to the blue line presented above, and using stay at home orders to replicate the orange line, policy makers can effectively replicate an integral feedback policy.}
    \label{fig:tv_betas}
\end{figure}

While there are many ways integral feedback can be implemented in practice, we highlight one possible policy which can be taken by adapting the existing policy in \eqref{eq:rho_def} using time-varying parameters.
In Figure \ref{fig:tv_betas}, we show that the integral policy suggested Figure \ref{fig:us_fit} can be replicated by allowing $\beta_1$ and $\beta_2$ to be time-varying.
Moreover, since we have shown that $\beta_1$ and $\beta_2$ relate linearly to observed policies such as vaccination rates and mobility above (Supplementary Information, \figpolicy), this suggests that using time-varying policies can recreate an integral feedback policy.
Specifically, in the United States we find that $\beta_1$ relates collinearly with vaccination rates, and that $\beta_2$ related collinearly with percentage of time individuals spend at home.
Using these policy levers in a way that is consistent with Figure \ref{fig:tv_betas} would effectively replicate an integral feedback policy over time.
Because the values of $\beta_1$ in Figure \ref{fig:tv_betas} are approximately ten times larger in magnitude than those learned across the 50 states, the analysis suggest that any appropriate policy would require that interventions would need to be increased by an order of magnitude in order to eradicate COVID-19 cases.


\section{Discussion}

We provide a dynamical model to represent the trajectory of cases in an epidemic, and show that COVID-19 cases are well explained by a control strategy which only depends on three parameters. 
The result is based on a robust estimation procedure, and all learned parameters are non-zero in a statistically significant sense. 

While the results are strong, we note that the simplicity of the model and its implicit nature do have some drawbacks, as it is difficult to capture the known properties of the disease in this model. 
For example, the assumption that cases in week $t+1$ are driven by the cases in week $t$ does not explicitly take into account possible delays in testing procedures and the amount of time it takes for an exposed individual to test positive.
However, since we do find that the model is sufficient to explain variation in the data and provides a rigorous data-driven approach to estimation by making such simplifying assumptions, we believe there is value in modeling epidemic dynamics using this approach.
In future work, we aim to better understand the determinants of this endogeneity, to understand why community responses to the pandemic can be easily explained by this low dimensional model. 
Such work would also be aimed at adapting this model for longer term predictions, as feedback is important in understanding long term trends of infectious disease \cite{rahmandad2020risk}.
As we move towards data-driven control of epidemics, our hope is to utilize this endogeneity so that policy can be designed optimally.

Ultimately, these results have important implications for the development of control strategies against infectious disease in the future. 
While the strategy that fits the data well does not result in an eradication of COVID-19 cases, augmenting the implicit control with a term proportional to the sum of previous case counts would eventually eradicate new cases.
This suggests that policy makers should emphasize the cumulative costs of the pandemic when communicating with the public, even when current cases are low.


\bibliographystyle{plain}
\bibliography{main.bib}

\appendix
\newpage
\noindent {\Large \textbf{Supplementary Information}}
\section{Proof of Proposition 1}
\label{a:proof}
\begin{proof}
First, note that for the two models to be equivalent, we must take the noise parameter in \branchingeq~to be a constant, i.e. $\sigma^2 = 0$.
We then construct the parameterization as follows: For arbitrary but particular time series $\beta(t)$ and $\gamma(t)$, first define
\[ \mathcal{R}^{SIR}(t) = 1 - \gamma(t) + \frac{\beta(t)}{N} S(t) \,.\]
Then, for
\begin{align*} 
\mathcal{R}(0) = \mathcal{R}^{SIR}(0)\,,\quad \text{and} \quad \forall t \geq 1, ~\rho(t) =\frac{\mathcal{R}^{SIR}(t)}{\mathcal{R}^{SIR}(t-1)} \,,~~ \,,
\end{align*}
the model in \branchingeq~is equivalent to the time-varying SIR model in \sireq.
\end{proof}

\section{Connection Between the Closed Loop Model and Time-Varying SIR Model}
\label{a:connection}
We note there is a connection between the SIR model in \sireq~ and the closed loop model in \closedloopeq~ in the sense that \rhodefeq~ implies a particular structure for the time-varying parameters in \sireq. Specifically, we see that from \rhodefeq,
\begin{equation}
    \rho(t) = I(t)^{\beta_1} \left( \frac{I(t)}{I(t-1)}\right)^{\beta_2} \mathcal{R}(0)^{-\beta_2} e^{\beta_3} \,.
\end{equation}
Furthermore, from the above proof of Proposition 1, we see that the open loop model and the time-varying SIR model are equivalent so long as
\begin{equation}
    \rho(t) = \frac{N - N \gamma(t) + \beta(t) S(t)}{N - N\gamma(t-1) + \beta(t-1) S(t-1)} \,.
\end{equation}
Equating these two representations of $\rho(t)$, we find that if $\gamma(t)$ and $\beta(t)$ obey
\begin{equation}
    \left(\frac{\gamma(t)}{\gamma(t-1)} \right) \left(\frac{\frac{N + \beta(t) S(t)}{\gamma(t)} - N}{\frac{N + \beta(t-1) S(t-1)}{\gamma(t-1)} - N} \right) = I(t)^{\beta_1} \left( \frac{I(t)}{I(t-1)}\right)^{\beta_2} \mathcal{R}(0)^{-\beta_2} e^{\beta_3} \,,
\end{equation}
which can be done by equating terms of the product with one another, then the two models can be equivalent.
For example, if for all $t \geq 1$,
\begin{equation}
    \frac{\gamma(t)}{\gamma(t-1)} = I(t)^{\beta_1} e^{\beta_3} \,, \quad \text{and}\quad \frac{\frac{N - \beta(t) S(t)}{\gamma(t)} - N}{\frac{N - \beta(t-1) S(t-1)}{\gamma(t-1)} - N} = \left( \frac{I(t)}{I(t-1)}\right)^{\beta_2} \mathcal{R}(0)^{-\beta_2} \,,
\end{equation}
then we see a specific choice of $\beta(t)$ and $\gamma(t)$ can be imposed by the closed loop model in \closedloopeq.
Specifically, this particular choice would result in
\begin{align*}
    \gamma (t) = \left(\prod_{k = 1}^{t-1} I(k) \right)^{\beta_1} e^{t \beta_3} \gamma (0) \,,
\end{align*}
and 
\begin{align*}
    \beta(t) = \left(N - \gamma(t) \left( \left( \frac{I(t)}{I(t-1)}\right)^{\beta_2} \mathcal{R}(0)^{-\beta_2} \left(\frac{N - \beta(t-1) S(t-1)}{\gamma(t-1)} - N \right) + N \right) \right) / S(t) \,.
\end{align*}
That is, one possible interpretation of the implicit feedback policy is that the magnitude of $\gamma(t)$ increases as the cumulative product of cases increases, and that $\beta(t)$ is inversely proportional to the remaining susceptible population and is a complex function of recovery rates $\gamma(t)$ and the rate of change of cases $I(t) / I(t-1)$.

\section{Additional Figures}
\label{a:additional_figs}
\begin{figure}[h]
    \centering
    \includegraphics[width=0.7\textwidth]{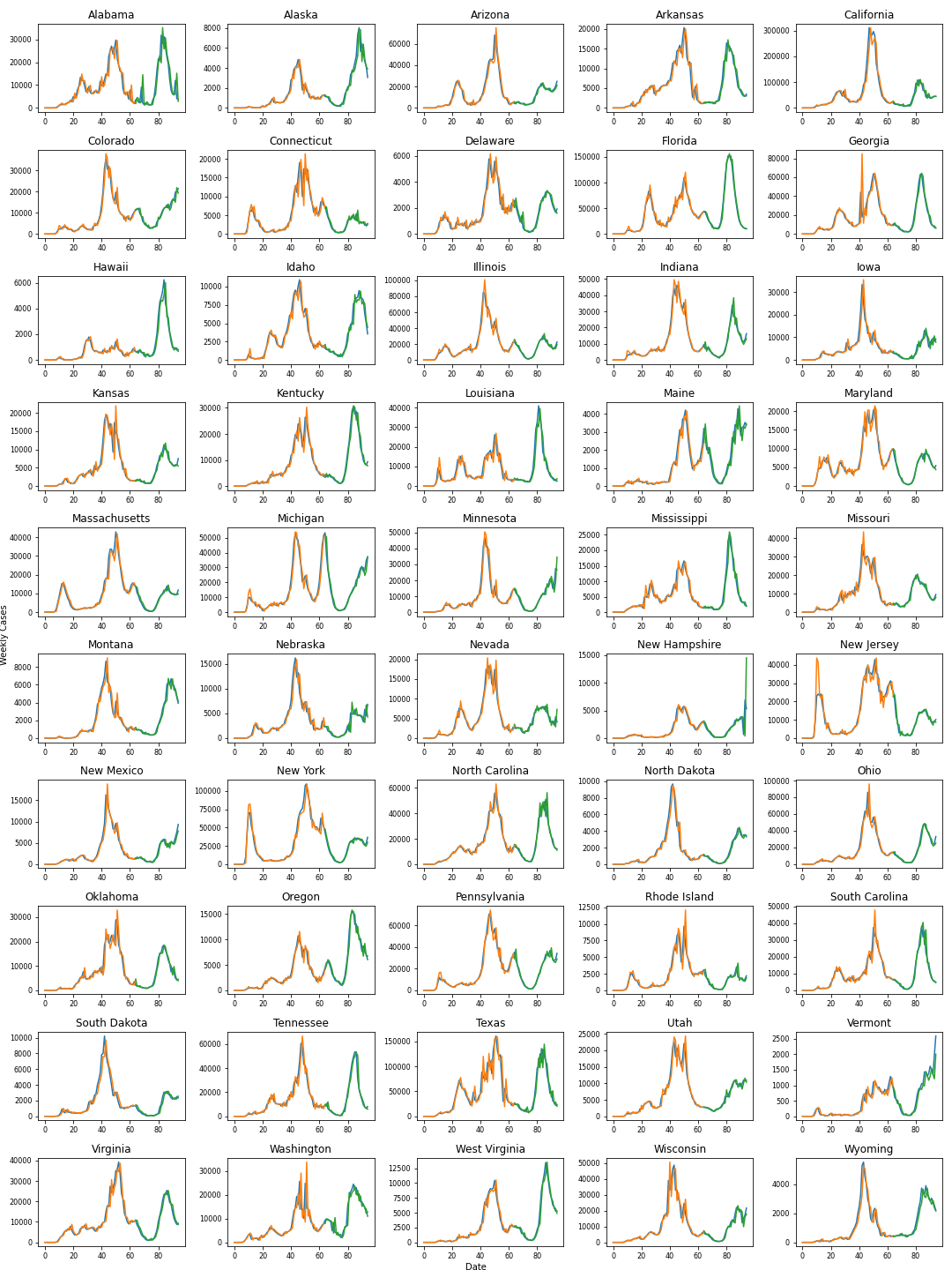}
    \caption{Case counts of US states as modeled by the dynamical system. One week ahead predictions using this method are shown in orange (in sample) and green (out of sample), and the true data is shown in blue. We find that using data from roughly the first year of the pandemic results in surprisingly accurate performance on later data.}
    \label{fig:state_fits}
\end{figure}

\begin{figure}[t]
    \centering
    \begin{subfigure}[b]{0.5\textwidth}
         \centering
         \includegraphics[width=0.9\textwidth]{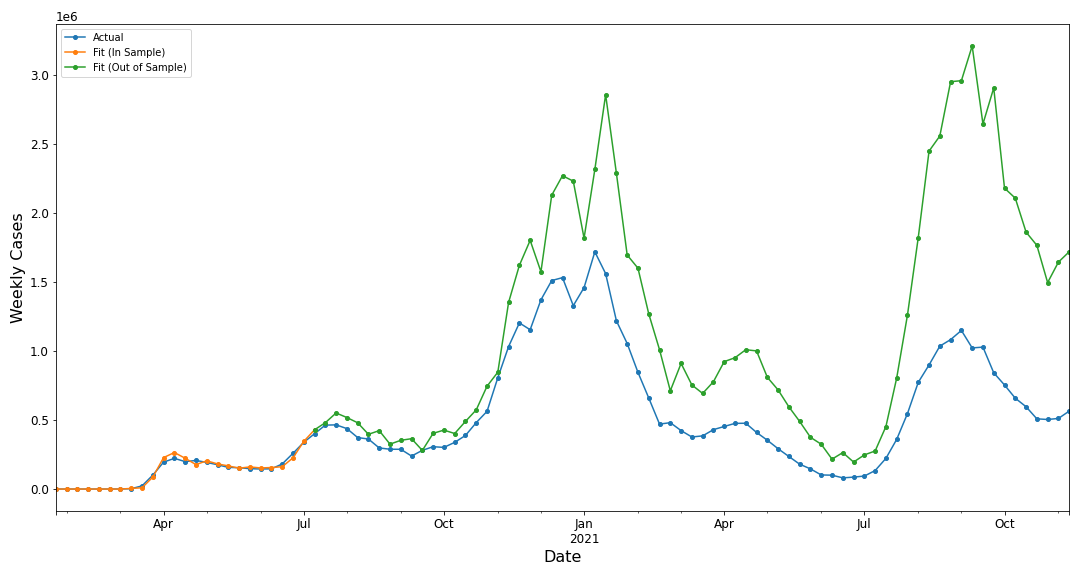}
         \caption{PID controller learned with data from the first wave of the COVID-19 pandemic in the United States.}
         \label{fig:pid_70}
     \end{subfigure}
     \begin{subfigure}[b]{0.5\textwidth}
         \centering
         \includegraphics[width=0.9\textwidth]{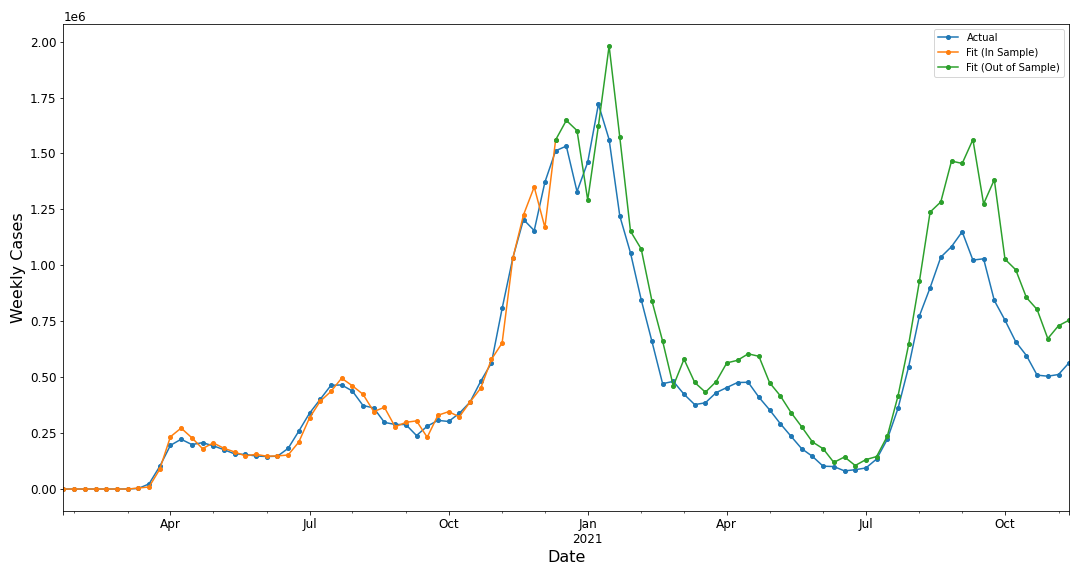}
         \caption{PID controller learned with data from the first two waves of the COVID-19 pandemic in the United States.}
         \label{fig:pid_48}
     \end{subfigure}
    \caption{Implementation of a PID controller learned using the single state $X(t) = \log I(t)$ (i.e., $X_2(t)$ in \closedloopeq). 
    The PID controller has the form $u(t) = k_P X(t) + k_I \sum_{\tau=1}^t X(\tau) + k_D (X(t) - X(t-1))$, and we assume the system has the form $X(t+1) = X(t) + u(t)$. 
    Learning for the first two waves results in an out of sample $R^2$ value of $0.085$, and from the first three waves results in $R^2 = 0.81$, compared to an $R^2$ of $0.905$ using the model in \closedloopeq~trained on the first two waves of data, implying that the PID method requires more data to provide a worse explanation out of sample.
    This suggests the model in \closedloopeq~provides a better fit to the data.}
    \label{fig:pid}
\end{figure}

\begin{figure}[t]
     \centering
     \begin{subfigure}[b]{0.5\textwidth}
         \centering
         \includegraphics[width=0.9\textwidth]{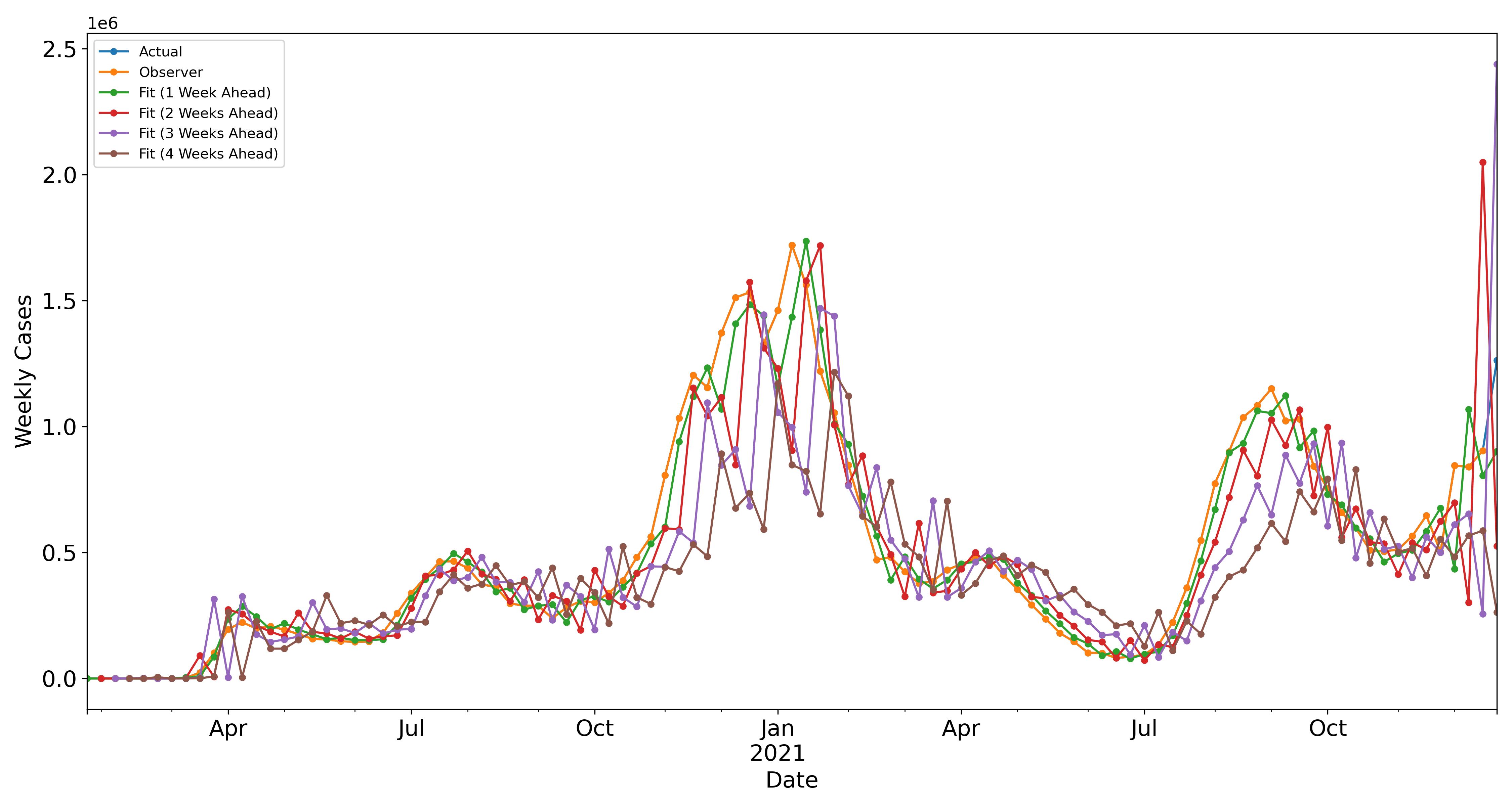}
         \caption{Predictions using statistical estimates.}
         \label{fig:stat}
     \end{subfigure}
     \begin{subfigure}[b]{0.5\textwidth}
         \centering
         \includegraphics[width=0.9\textwidth]{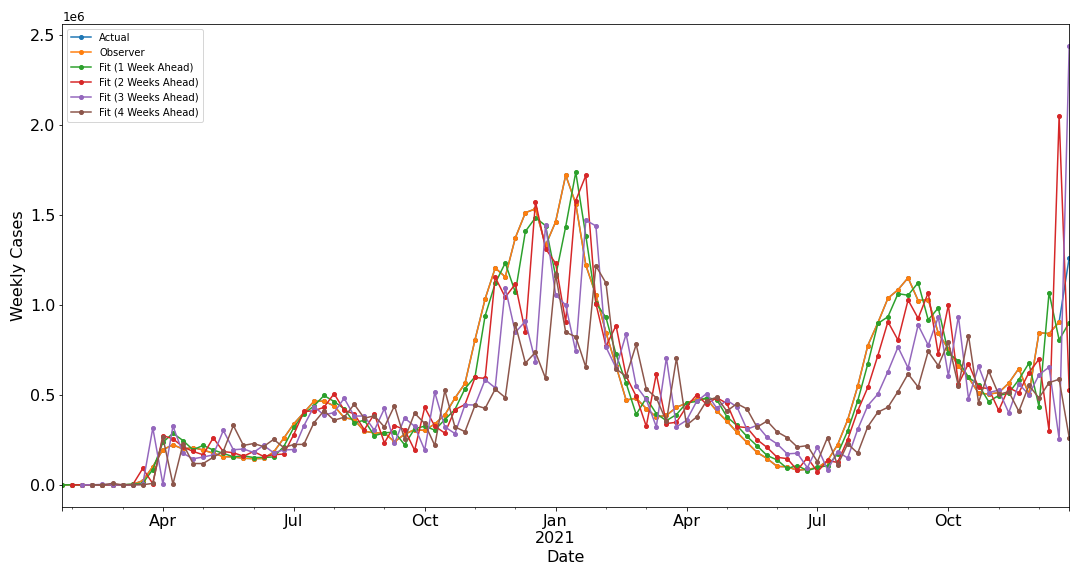}
         \caption{Predictions using estimates from an observer system.}
         \label{fig:observer}
     \end{subfigure}
     \begin{table}[H]
         \centering
         \begin{tabular}{|l|r|r|r|r|} \hline
         $k$ & 1 & 2 & 3 & 4 \\ \hline
            Statistical Estimates & 0.949 & 0.607 & 0.267 & -1.562 \\ \hline
            Observer System  & 0.949 & 0.607 & 0.267 & -1.562 \\ \hline
            CDC Baseline & 0.942 & 0.861 & 0.765 & 0.648 \\ \hline
         \end{tabular}
         \caption{$r^2$ Comparison of $k$-week ahead forecasts}
         \label{tab:my_label}
     \end{table}
        \caption{$k$-week ahead forecasts using the system in \closedloopeq~using statistical estimates (a) or an observer model (b) which uses an estimate of $X_1(t)$. For statistical estimates, future values of $X_1(t)$ are estimated through feeding back estimates into \closedloopeq, and for the observer system a parallel system is used.
        We find that using the observer system results in the same performance compared to the statistical method, and that the closed loop method works best in the short term for 1 to 2 week ahead predictions.}
        \label{fig:k_week_ahead}
\end{figure}

\begin{figure}[t]
    \centering
    \includegraphics[width=0.8\textwidth]{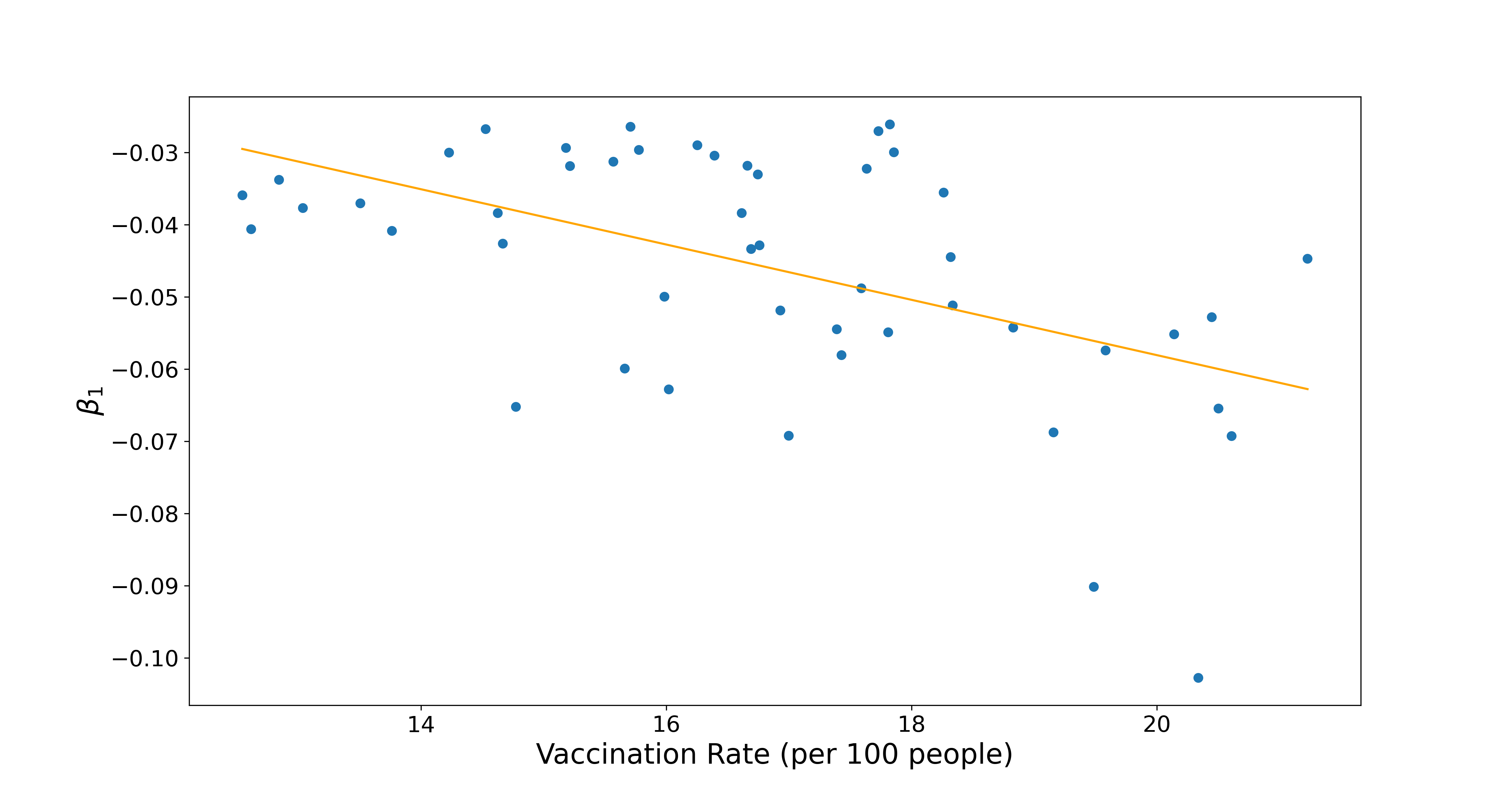}
    \caption{Comparison of learned $\beta_1$ parameters in each state to observed vaccination rates from \url{https://github.com/owid/covid-19-data}, where each point represents a state and the orange line is the line of best fit.
    We see that higher vaccination rates correspond to larger magnitudes of the $\beta_1$ parameter.
    }
    \label{fig:param_comparison}
\end{figure}

\begin{figure}[t]
    \centering
    \includegraphics[width=0.8\textwidth]{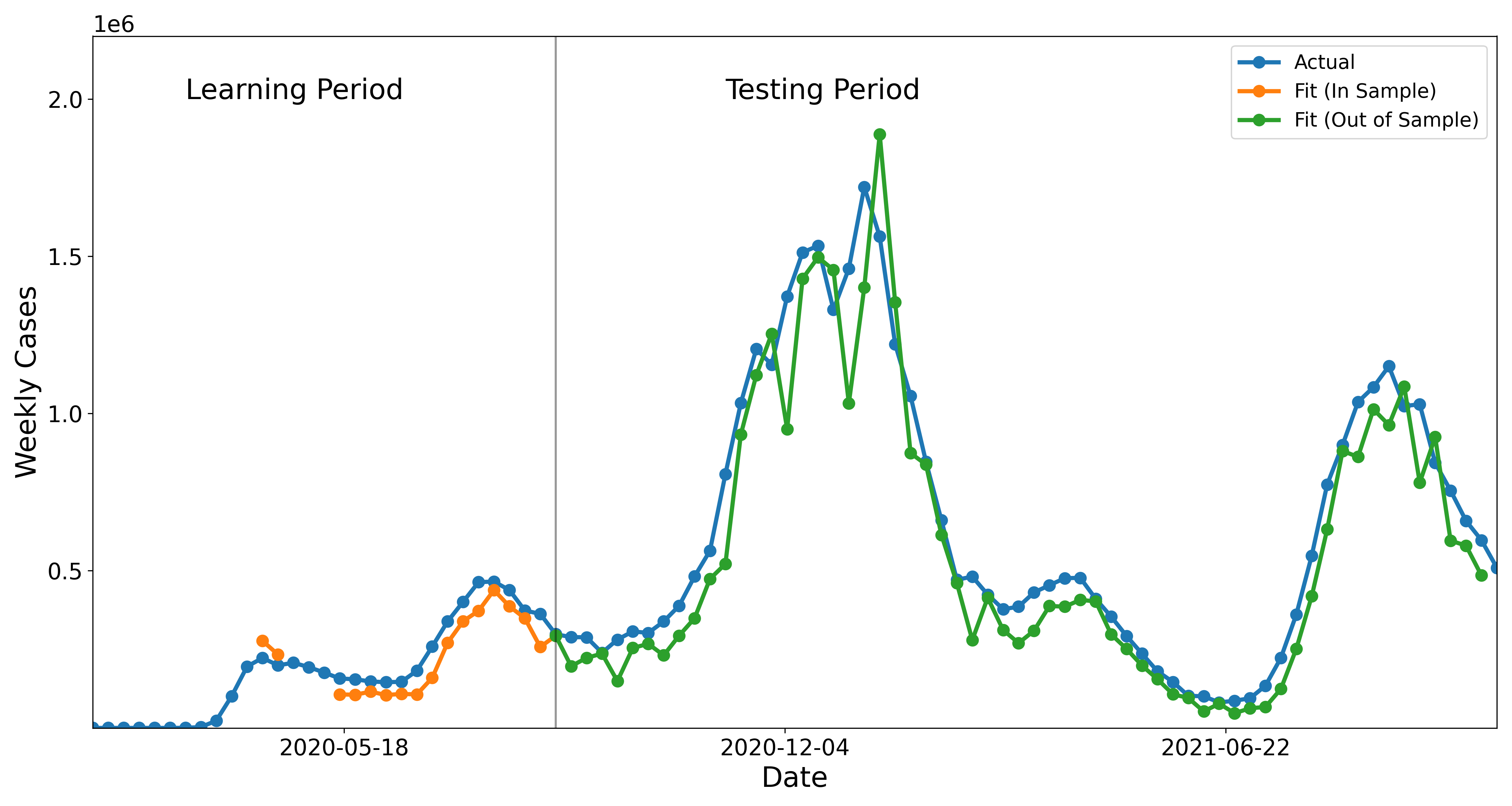}
    \caption{Forecasts of United States COVID-19 Cases using data available at each date.
    Because case data is adjusted after the fact, we find that using data available at each date results in a small deterioration in prediction performance, with an out of sample $R^2$ score of $0.901$ as opposed to $0.948$ using data as of March 2022.}
    \label{fig:as_of}
\end{figure}

\begin{table*}[t]
\centering
\begin{tabular}{lrrr}
\toprule
State &  $R^2$ of Implicit Feedback Predictions&    $R^2$ of CDC Forecasts & $R^2$ of SEIR Model\\
\midrule
Alabama        &   0.761268 &  0.870218 &  -7.466713 \\
Alaska         &   0.895081 &  0.860655 &  -6.097458 \\
Arizona        &   0.948373 &  0.912740 & -16.846574 \\
Arkansas       &   0.919352 &  0.904562 &  -5.102133 \\
California     &   0.846362 &  0.815874 & -38.857268 \\
Colorado       &   0.912670 &  0.928491 & -17.163285 \\
Connecticut    &   0.787693 &  0.770680 & -13.130171 \\
Delaware       &   0.901341 &  0.960540 & -23.524277 \\
Florida        &   0.984187 &  0.934497 & -16.086748 \\
Georgia        &   0.967953 &  0.951781 & -29.483336 \\
Hawaii         &   0.910708 &  0.919715 &  -3.725094 \\
Idaho          &   0.936136 &  0.919588 & -86.226929 \\
Illinois       &   0.923297 &  0.902908 & -25.749134 \\
Indiana        &   0.902190 &  0.925885 & -61.444810 \\
Iowa           &   0.813681 &  0.777314 &  -8.914433 \\
Kansas         &   0.930767 &  0.851023 &  -8.554606 \\
Kentucky       &   0.961193 &  0.927413 &  -4.442309 \\
Louisiana      &   0.895427 &  0.894736 & -25.563110 \\
Maine          &   0.854193 &  0.871618 & -13.050083 \\
Maryland       &   0.941903 &  0.949353 & -29.439844 \\
Massachusetts  &   0.920278 &  0.905620 & -12.569417 \\
Michigan       &   0.937189 &  0.953322 & -20.089249 \\
Minnesota      &   0.876778 &  0.852998 &  -8.951495 \\
Mississippi    &   0.931766 &  0.887371 &  -5.910868 \\
Missouri       &   0.952072 &  0.903154 & -12.156490 \\
Montana        &   0.956009 &  0.956591 &  -3.621835 \\
Nebraska       &   0.648975 &  0.606238 & -16.876655 \\
Nevada         &   0.786819 &  0.794385 & -32.207349 \\
New Hampshire  &   0.337766 &  0.012839 & -32.442966 \\
New Jersey     &   0.913044 &  0.947180 &  -7.806754 \\
New Mexico     &   0.918922 &  0.858679 & -12.438151 \\
New York       &   0.948178 &  0.925340 &  -8.925778 \\
North Carolina &   0.892444 &  0.908425 &  -7.784810 \\
North Dakota   &   0.969735 &  0.964937 &  -3.343721 \\
Ohio           &   0.965163 &  0.950134 & -19.912066 \\
Oklahoma       &   0.957803 &  0.922573 &  -7.961322 \\
Oregon         &   0.949682 &  0.935502 &  -2.391805 \\
Pennsylvania   &   0.937376 &  0.950664 & -20.954253 \\
Rhode Island   &   0.756697 &  0.821439 & -17.282322 \\
South Carolina &   0.905781 &  0.915286 &  -9.553737 \\
South Dakota   &   0.929809 &  0.936414 & -13.868844 \\
Tennessee      &   0.935708 &  0.920788 & -41.321343 \\
Texas          &   0.884807 &  0.915912 & -10.169516 \\
Utah           &   0.945423 &  0.865068 & -16.726585 \\
Vermont        &   0.832917 &  0.819457 & -45.831707 \\
Virginia       &   0.956532 &  0.954946 & -16.765430 \\
Washington     &   0.847929 &  0.893503 &  -5.760364 \\
West Virginia  &   0.957164 &  0.947691 &  -2.813714 \\
Wisconsin      &   0.897933 &  0.870112 & -12.464855 \\
Wyoming        &   0.924037 &  0.919302 &  -9.180029 \\
\midrule
\textbf{Mean} & 0.891370 & 0.877309\\
\bottomrule
\end{tabular}
    \caption{$r^2$ metric of 1 week ahead forecasts using the implict feedback approach summarized in \closedloopeq~compared to the $r^2$ scores of 1 week ahead forecasts from a state of the art ensemble model \cite{ray2020ensemble} and a simple SEIR model \cite{hethcote2000mathematics}.
    }
    \label{tab:state_r2}
\end{table*}

\begin{table*}[t]
    \centering
    \begin{tabular}{clrrc}
\toprule
Dependent Variable &   Independent Variable &  Coefficient Value &  $p$-value & \shortstack{Null Hypothesis\\ Rejected} \\
\midrule
         $\beta_2$ &   Mobility Component 2 &              0.006 &      $<0.001$ &                            $\checkmark$ \\
         $\beta_2$ &  Natural Immunity (\%) &             16.852 &      $<0.001$ &                            $\checkmark$ \\
         $\beta_2$ &   Mobility Component 1 &              0.002 &      $<0.001$ &                            $\checkmark$ \\
         $\beta_1$ &   Mobility Component 2 &              0.001 &      0.001 &                            $\checkmark$ \\
         $\beta_1$ &  Natural Immunity (\%) &              5.233 &      0.001 &                            $\checkmark$ \\
         $\beta_1$ &   Mobility Component 1 &              0.000 &      0.443 &                                         \\
\bottomrule
\end{tabular}
    \caption{Hypothesis tests comparing different policies to learned $\beta_1$ and $\beta_2$ values across both US states and countries around the globe. 
    Across heterogeneous regions, we find that $\beta_1$ and $\beta_2$ tend to correlate significantly with observed measures such as mobility and the level of natural immunity.}
    \label{tab:policy_corr}
\end{table*}

\begin{table*}[t]
    \centering
\begin{tabular}{clrrc}
\toprule
Dependent Variable &              Independent Variable &  Coefficient Value &  $p$-value & \shortstack{Null Hypothesis\\ Rejected} \\
\midrule
         $\beta_1$ &             Vaccination Rate (\%) &             -0.004 &      $<0.001$ &             $\checkmark$ \\
         $\beta_1$ &              Mobility Component 2 &             -0.000 &      0.004 &             $\checkmark$ \\
         $\beta_1$ &  Mean Distance Traveled From Home &             -0.028 &      0.007 &             $\checkmark$ \\
         $\beta_2$ &       Median Percentage Time Home &              0.016 &      0.018 &             $\checkmark$ \\
         $\beta_2$ &  Mean Distance Traveled From Home &             -0.210 &      0.044 &                          \\
         $\beta_2$ &              Mobility Component 1 &             -0.001 &      0.056 &                          \\
         $\beta_1$ &  Median Time Spent Away From Home &              0.000 &      0.064 &                          \\
         $\beta_2$ &             Vaccination Rate (\%) &             -0.019 &      0.073 &                          \\
         $\beta_1$ &                      Mask Mandate &             -0.017 &      0.079 &                          \\
         $\beta_2$ &             Natural Immunity (\%) &              7.116 &      0.128 &                          \\
         $\beta_2$ &              Mobility Component 2 &              0.002 &      0.184 &                          \\
         $\beta_1$ &             Natural Immunity (\%) &              0.608 &      0.210 &                          \\
         $\beta_2$ &                      Mask Mandate &             -0.057 &      0.550 &                          \\
         $\beta_2$ &  Median Time Spent Away From Home &             -0.001 &      0.582 &                          \\
         $\beta_1$ &       Median Percentage Time Home &             -0.000 &      0.853 &                          \\
         $\beta_1$ &              Mobility Component 1 &             -0.000 &      0.962 &                          \\
\bottomrule
\end{tabular}
    \caption{Hypothesis tests comparing different policies to learned $\beta_1$ and $\beta_2$ values in the United States only. We find that $\beta_1$ correlates with vaccination rates in each state as well as mobility, and that $\beta_2$ correlates with mobility metrics.
    Here, vaccination data is taken from \url{https://github.com/owid/covid-19-data} and mobility auxiliary mobility data is taken from SafeGraph (``Median Time Spent Away From Home,'' ``Mean Distance Traveled From Home,'' , ``Median Percentage Time Home'') and Google (Mobility Components 1 and 2, which are principal components of the 6 mobility metrics provided by Google).}
    \label{tab:policy_corr_state}
\end{table*}

\end{document}